\documentclass[twocolumn,showkeys,preprintnumbers,amsmath,amssymb]{revtex4}
\usepackage{graphicx}
\usepackage{bm}
\usepackage{graphicx}

\usepackage{epstopdf}
\usepackage{psfrag}  
\usepackage{amsmath,amsthm}
\usepackage{bm}

\begin{document}

\preprint{APS}

\title{Exploring the randomness of Directed Acyclic Networks}
\author{Joaqu\'in Go\~ni$^{*,2}$, Bernat Corominas-Murtra$^{*,1}$, Ricard V. Sol\'e$^{1,3,4}$ and Carlos Rodr\'iguez-Caso$^1$}

\affiliation{$^*$ both authors contributed equally to this work\\
$^1$ ICREA-Complex Systems  Lab, Universitat Pompeu Fabra
  (PRBB). Dr Aiguader 88, 08003 Barcelona, Spain\\ 
$^2$Functional
  Neuroimaging  Lab. Department of  Neurosciences. Center  for Applied
  Medical Research. University of Navarra. Pamplona, Spain\\ 
$^3$Santa
  Fe Institute, 1399 Hyde Park Road, New Mexico 87501, USA\\
$^4$Institut de Biologia Evolutiva. CSIC-UPF. Passeig Maritim de la Barceloneta, 37-49, 08003 Barcelona, Spain.}

\begin{abstract}
The feed-forward relationship naturally observed in time-dependent
processes  and  in  a diverse  number  of  real  systems -such  as  some
food-webs and electronic and neural  wiring- can be described in terms
of so-called directed  acyclic graphs (DAGs).  An important ingredient
of  the analysis  of such  networks is  a proper  comparison  of their
observed  architecture against an  ensemble of  randomized graphs,
thereby  quantifying the  {\em randomness}  of the  real  systems with
respect to suitable  null models.   This  approximation is  particularly
relevant  when  the finite  size  and/or  large  connectivity of  real
systems  make inadequate  a comparison  with the  predictions obtained
from  the  so-called {\em  configuration  model}.   In  this paper  we
analyze four  methods of DAG randomization  as defined by
the  desired  combination  of  topological  invariants  (directed  and
undirected degree  sequence and  component distributions) aimed  to be
preserved. 
A highly ordered DAG, called \textit{snake}-graph and a Erd\:os-R\'enyi DAG were used to validate the performance
of   the  algorithms. Finally, three  real case studies, namely,  the  \textit{C.   elegans}  cell  lineage
network,  a PhD  student-advisor  network and  the Milgram's  citation
network were  analyzed using each  randomization method.   Results  show   how  the   interpretation  of
degree-degree relations in DAGs  respect to their randomized ensembles
depend on  the topological invariants  imposed. In general,  real DAGs
provide disordered  values, lower  than the expected  by chance  when the
directedness of the links is not preserved in the randomization process.
Conversely, if the direction of  the links is conserved throughout the
randomization process,  disorder indicators are close  to the obtained
from the null-model ensemble, although some deviations are observed.


\end{abstract}


\keywords{complex  networks, directed  acyclic graphs,  random models,
  graph randomization, null models}

\maketitle

\section{Introduction}

Many relevant  properties of  complex systems can  be described  by an
appropriate network representation  of their elements and interactions
\cite{Watts1998,      Newmann2001,     Dorogovtsev2003,     Sole:2003,
  Boccaletti2006}. Most of these  networks are  directed, i.   e.  there  is a
directional relationship between  two elements defining who influences
who in a given order. Among  the  class  of   directed  networks,  directed  acyclic  graphs
-henceforth, DAGs-  are an important  subset lacking feedback  loops.  This is specially suitable for
the  representation  of  evolutionary,  developmental  and  historical
\textit{processes}   in  which   the  time   asymmetry   determines  a
feed-forward (acyclic) flow  of causal relations. In this context, DAGs
constitute  a formal  representation  of causal  relations  that display  the
direct effects  of earlier events over latter  ones. Citation networks
are among their most paradigmatic  cases \cite{Hummon1989, Gardfield1964}.  In these networks nodes are
scientific   articles  and   directed  links   (or  arcs)   stand  for
bibliographic  citations  among them.   According  to a  chronological
order, directed  links are established  from former articles  to newer
ones in  a feed-forward manner.

In  general, time-dependent  processes have  been formalized  as DAGs. Examples of that comprehend
article and patent citation networks \cite{Valverde2007, Csardi2007}, decision jurisprudence  processes \cite{Fowler2008, Chandler2005} and tree genealogies
and  phylogenies.  Moreover,  other relevant  systems such  as standard
electric   circuits  \cite{Clay1971},   feed-forward   neural
\cite{Haykin1999}  and   transmission  networks  \cite{Frank1972}  are also
suitably represented as DAGs.

The main  objective of this paper  is to explore  the {\em randomness}
-in topological  terms- of real systems displaying  a directed acyclic
structure  by the defintion a collection of randomization methods  that preserves  a fixed
number of  topological invariants.   To this end,  the design  of null
models to  highlight the  particular features characterizing  a system
with respect to a  neutral or {\em random} scenario \cite{Newmann2001}
is needed.
In this context, the   so-called    \textit{configuration   model}   \cite{Newmann2001,
  Dorogovtsev2003,  bollobas01,  Molloy1995}  has  been  probed  as  a
fruitful  approximation to provide  a null-model  scenario of  what is
expected by \textit{chance} in  complex networks under the assumptions
of sparseness,  infinite size and lack of correlations.   However,
little attention  has been paid  concerning DAGs. Indeed, a
rigorous definition  of random DAG  from its directed  degree sequence
has been recently proposed  \cite{Karrer2009}, rising the interest for
its  study  through   configuration  model  approach.   Borrowing  the
methodology to build  random undirected  graphs \cite{Molloy1995,
  Aiello:2001}, degree sequence is  visualized as a
set of edge \textit{stubs}. Hence a random DAG is constructed by matching
stubs according  to certain  order constraints  until  they are
completely  canceled   \cite{Karrer2009}.   Without  neglecting  the
important  advance  it  represents  for the  comprehension  of  acyclic
networks, some  problems arise in  using this methodology as  the null
model reference  of real nets. Firstly, this  methodology is dependent
on how probable is to construct  a graph from a degree sequence, since
not all  of them produce  a graph, i.e., they are not \textit{graphical}.  
Additionally,  configuration model  assumptions are not fulfilled in real
systems due to their finite size and the  presence of  densely connected
regions.

An alternative approach  used in this work is  based on iterative processes
of  edge  rewiring  over  the  graph,  keeping  the  \textit{graphical}
condition during all the process  of randomization. This is a relevant
issue since the degree sequence, either directed or undirected, imposes a particular space of topological
configurations  rather limited  -as we  shall  see in  this work-  for
DAGs. Attending to this approach we can estimate where a real graph is
placed preserving a graphical ensemble that holds some topological
invariants.
The two fundamental topological invariants considered in this work for
a null  model comparison are  the degree sequence (either  directed or
undirected)  and the  component  structure.  The  both types of degree sequences  and the
degree distribution  have been typically chosen  as invariant in  the random
model   construction   \cite{Molloy1995,   Newmann2001,   Maslov2002a,
  Karrer2009}.  However, as is well  known in random graph theory, the
existence of  some graph satisfying  a given degree sequence  does not
guarantee  a single connected  component containing  the whole  set of
nodes,  except  at  high  connectivities.  Therefore  sparse  networks
representing connected systems are  expected to be fragmented during a
randomization  process.   This may  be  an  undesirable effect  when
studying historical  processes since it breaks the  flow of causality.
Besides this  problem, there are  also real systems that  display more
than a  single connected component.  Those  disconnected components do
not  interact  among them  and,  arguably,  can  be considered  to  be
independent systems  in terms of  causality.  It is worth to note that preservation  
of   connected   component  in   graph randomization processes
  has recently raised the interest  of network community \cite{Hanhijarvi09}.

According to the above  considerations, in order to produce comparable
ensembles for the evaluation of the randomness of a real DAG, we propose a
collection  of four  randomization  methods for  DAGs  that ensure  the
topological   invariants   mentioned   above.  These
randomization  techniques  were applied  to two extreme  -in terms of degree-degree relations- network models:  an   Erd\"os  R\'enyi  DAG   and  a  highly
ordered graph called \textit{snake}-DAG.  Our methodology was then applied
to three real DAGs: a  citation network, a PhD-student advisor network
and  the cell  lineage in  the development  of \textit{caebnorhabditis
  elegans} worm.

The  paper  is organized  as  follows:  Section  II offers  the  basic
concepts related to  DAGs.  Section III explicitly defines  the set of
four  randomization  algorithms   according  to  different  topological
invariants.   Section IV  describes and  characterizes  the randomness
indicators and  apply the randomization processes to  the systems under
study: two toy models -which  enable us to validate the performance of
the  algorithms-  and three  real  systems.  Section  V discusses  the
relevance of the obtained results.

\section{Ordering, causality and formal definition of DAGs}

In this  section we discuss  some mathematical properties of  DAGs and
their interpretation in terms  of causal relations. We finally address the problem of component structure conservation.

\subsection{Basic definitions}

Let ${\cal  G}(V, E)$ be  a directed graph,  being $V=\{v_1,...,v_N\}$
the  set of  nodes, and  the set  of ordered  pairs  $E=\{\langle v_k,
v_i\rangle, ...,  \langle v_j, v_l\rangle\}$  the set of  edges -where
the order, $\langle v_k, v_i\rangle$ implies that there is an arrow in
the  following direction: $v_k\rightarrow  v_i$.  The  {\em underlying
  graph}  ${\cal G}_u$ of a  directed graph ${\cal G}$ is
an undirected graph  with the same set of  nodes ${\cal G}$, but
whose  edges are undirected  (i.e.  $\langle  v_k, v_j\rangle  \in E$,
then  $\{\langle   v_j,v_k\rangle,\langle  v_k,v_j\rangle\}\in  E_u$).
Given a  node $v_i\in V$,  the number of  outgoing links, to  be noted
$k_{o}(v_i)$, is called the {\em  out-degree} of $v_i$. Similarly, the number
of ingoing links of $v_i$ is called {\em in-degree} of $v_i$, noted by
$k_{i}(v_i)$.

A DAG is  a directed graph characterized by the  absence of cycles: If
there is a  {\em directed path} from $v_i$ to $v_k$  (i.e., there is a
finite   sequence   $\langle  v_i,   v_j\rangle,   \langle  v_j,
v_l\rangle,\langle v_l,  v_s\rangle, ..., \langle  v_m, v_k\rangle \in
E$)  then,  there  is  no directed path  from  $v_k$  to  $v_i$.
Borrowing  concepts from order  theory \cite{Suppes:1960, Kelley:1955}, we  refer to  nodes with
$k_{i}=0$ as {\em maximals}  and those with $k_{o}=0$ as {\em
  minimals}.  The absence of cycles ensures that at least there is one
minimal node and one maximal  node.  Maximal nodes  can be seen as  inputs of a  given computational  or sequential
process  while  minimal -or terminal-  ones are the  outputs of  such a  process.
Furthermore the acyclic nature permits to define  a node ordering
by labeling all nodes with sequential natural numbers.  Thus, in a
DAG there is at least one numbering of the nodes such that:
\begin{equation}
(\forall \langle v_i, v_j \rangle \in E)\Rightarrow(i>j).
\label{i>j}
\end{equation}
For this reason, DAGs have  been also referred as {\em ordered graphs}
\cite{Karrer2009}.  

\subsection{Random DAGs}

The theoretical  roots of the concept  of a random DAG are  based on the
so-called  directed degree  sequence  \cite{Karrer2009} -as  well as  the
concept of random graph \cite{Molloy1995}.  A random DAG ${\cal G}$ is
a  randomly chosen  element of  an ensemble  of DAGs  which  share the
directed degree  sequence, denoted by  $d({\cal G})$,  which is  defined as
follows:
\begin{equation}
d({\cal G})=(k_{i}(v_1), k_{o}(v_1)), ..., (k_{i}(v_i), k_{o}(v_i)),... 
\label{degreeSeq}
\end{equation}
The  two  numerical  quantities  composing  every element  of  such  a
sequence, $k_{i}(v_k)$ and $k_{o}(v_k)$,  encode the pattern
of connectivity of every node  of the graph.  In general, the ensemble
of  random graphs  containing $N$  nodes is  composed by  all possible
graphs whose connectivity pattern satisfies the directed degree sequence. If we
only pay  attention to the number  of edges connected to  a given node
$v_i$  -regardless the  direction of  the arrows-  we define  the {\em
  degree}         of         the         node         $v_i$         as
$k(v_i)=k_{i}(v_i)+k_{o}(v_i)$  \footnote{Such   equality  is  only
  general in DAGs, since the absence of cycles avoids the existence of
  autoloops  or  situations  like  $\langle v_i,  v_k\rangle,  \langle
  v_k,v_i\rangle\in  E$.  Furthermore,  within this  formalism,  it is
  assumed that we can neglect  the probability that two links begin at
  a  given node  $v_k$  and end  in a  given  node $v_i$,  due to  the
  assumption of {\em sparseness}. As  we shall see in section IV, such
  an  assumption   does  not  hold  for  some   real  systems.}   and,
consistently,  the {\em  undirected  degree sequence}  of ${\cal  G}$,
is defined as:
\begin{equation}
d_u({\cal G})=k(v_1), ..., k(v_i),... 
\label{degreeuSeq}
\end{equation}
However, it  is clear that  not any sequence  of $N$ pairs  of natural
numbers -or $N$  natural numbers in the case  of the undirected degree
sequence- represents the degree sequence of an ensemble of some kind
of  random graphs  containing $N$  nodes \cite{Molloy1995,Karrer2009}.
There are several restrictions that  a (un)-directed degree sequence must satisfy in
order to represent  a proper graph, i.e.,  a sequence
to  be {\em graphical}  or {\em  feasible} \cite{Molloy1995}.  In  the case  of directed
graphs, {\em in}  and {\em out} degrees of the  whole sequence must be
consistent with the number of edges, i.e.:
\begin{equation}
\sum_{i\leq n}k_{i}(v_i)=\sum_{i\leq n}k_{o}(v_i)=|E|,
\label{cond1}
\end{equation}
It  is clear  that such  a condition  does not  avoid the  presence of
cycles  in the network  structure.  Consistently  with the  claim that
DAGs depict systems where some  unavoidable ordering among nodes is at
work, we can ensure the generation of a given DAG if and only if there
is  a  labeling of  the  nodes such  that  $v_i\to  v_j$ implies  that
$i>j$\footnote{We observe  that we defined  an ordering which  is just
  the opposite  than the one defined in  \cite{Karrer2009}. The reason
  for this stems  from the interesting role played  by order theory to
  understand the  particular properties of  DAGs. In this way,  in our
  ordering, a  {\em maximal} will display  a label higher  than any of
  its  neighbors,  and  the  opposite  happens in  the  case  of  {\em
    minimal}; leading  this definition of  order to be  more intuitive
  for the reader. It is clear, however, that any choice is equivalent,
  provided that the construct  is internally consistent.}.  Taking into account this ordering  to build the graph, the directed degree sequence
must also hold two conditions. First,
\begin{equation}
k_{o}(v_1)=k_{i}(v_n)=0,
\label{cond2}
\end{equation}
and second,
\begin{equation}
(\forall v_i\in V)\;\;\sum_{j< i}k_{i}(v_j)-\sum_{j\leq i}k_{o}(v_j)\geq 0.
\label{cond3}
\end{equation}
Under  conditions (\ref{cond1},\ref{cond2},\ref{cond3}) it is ensured
that a directed degree sequence  will be graphical  and able to  represent the
degree sequence of a given non-empty ensemble of DAGs.

\subsection{Component structure and causal relations}

The concept  of component structure stems from the  notion of  undirected path:
given two  pairs of nodes  $v_i, v_k\in V$,  there is an  {\em undirected
  path} among them  if there is a finite  sequence of undirected edges
such that it can be ordered  sequentially, for example $\{ v_i, v_j\}, \{
v_j,  v_l\},\{  v_l, v_s\},  ...,  \{ v_m,  v_k\}  \in  E_u$.  A  {\em
  component} of  ${\cal G}$ is a  (maximal) subset of $V$  by which an
undirected path  can be defined among  any pair of  nodes. The special
features  of a  DAG impose  constraints on  the number  of  -DAG like-
components  from  a  given directed degree  sequence.   Indeed,  let  $M({\cal
  G})=\{v_k\in V:  k_{i}(v_k)=0\}$ be the  set of maximal nodes  of a
given DAG ${\cal G}$  and $\mu({\cal G})=\{v_k\in V: k_{o}(v_k)=0\}$
the set of its minimal nodes. Let $d({\cal G})$ be the directed degree sequence
of ${\cal G}$ which, by assumption, is graphical.  Then, the number of
(DAG) components of the graph $c({\cal G})$ is bounded by
\begin{equation}
c({\cal G})\leq \min\{|M|,|\mu|\},
\label{boundary}
\end{equation}
since any  connected DAG must  have, at least,  $|\mu|=|M|=1$.  Another
constraint  must  be  satisfied.  There  must  exist  a
partition of the directed degree sequence  $d({\cal G})$ by which all $c ({\cal
  G})$  subsequences  are  graphical,  i.e.,  they  satisfy  equations
(\ref{cond1},\ref{cond2},\ref{cond3}).

\section{Randomization Methods}

\begin{figure}
\begin{center}								
\includegraphics[width=8cm]{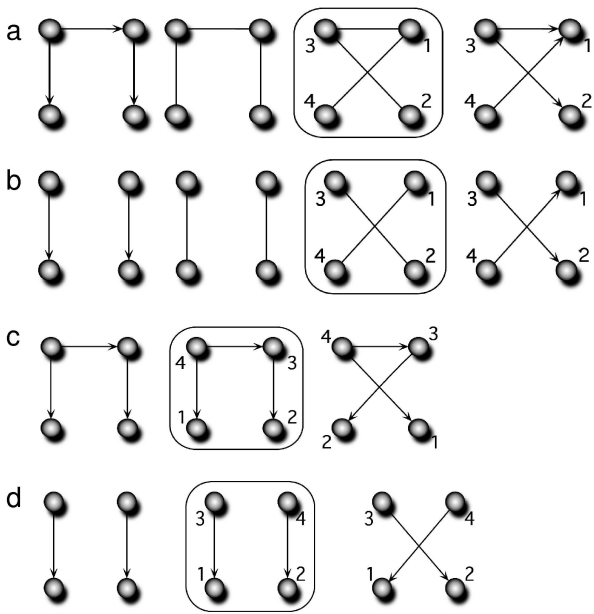}			%
 \caption{Schematic   representation   of    the   four   different   DAG
   randomizations  proposed in  this  work. Methods are  alphabetically denoted  (a to d). Method  $\mathbf{a}$: randomization  preserving
   undirected   degree  sequence   and  component   size  distribution. Method  $\mathbf{b}$: DAG  randomization  only preserving  the  undirected
   degree  sequence. Method  $\mathbf{c}$: randomization  preserving  directed
   degree  sequence  and  component  size  distribution.
   Method  $\mathbf{d}$: randomization    preserving   only    directed    degree   sequence.}
\label{random_method}	
\end{center}								%
\end{figure}

This section describes  four methods to obtain randomized ensambles of DAGs,
preserving  a number  of invariants.
The  four   algorithms  presented  below   perform  random  rewirings
according to constraints based  on (un)-directed degree sequence and component structures, all
of them  explicitly avoiding the presence  of multi-edges.  Algorithms
presented    in    this     work    are    illustrated    in    figure
\ref{random_method}. For the sake of clarity the order of presentation
of the  methods is the same for  all figures and tables,  and thus the
letters used  in fig. (\ref{random_method})  inequivocaly identify the
methods of randomization.

\subsection{Generating the ensemble from the undirected degree sequence}

The simplest method of randomization preserving components consists of
applying  a  random
numbering to ${\cal G}^u$.  This allows us to define  an order  criteria  to  establish  the
direction of arrows.   In this case, given an  undirected pair $\{v_i,
v_j\}$ we say that if $i>j$ we defined the order pair as $\langle v_j,
v_i\rangle$,  otherwise $\langle v_i,  v_j\rangle$. Since
${\cal G}^u$ is preserved, undirected degree-degree relations are also conserved.  Then,  a suitable randomization
requires  some additional  process  of link  rewiring  to destroy  the
presence of  degree-degree relations.  In  this context, we  provide a
methodology  that combines  a rewiring  process that  preserves undirected degree
sequence  -see  eq.  (\ref{degreeuSeq})-,  component  structure and
renumbering.

This first randomization method, denoted by the letter ${\mathbf { a}}$, is
depicted in fig.  (\ref{random_method}a).   The steps of the algorithm
are scheduled in the following:
\begin{enumerate}
\item
Given  a DAG,  ${\cal G}(V,  E)$, we  obtain its  respective underlying
network, ${\cal G}_u$.
\item
We obtain  a random network  conserving the undirected degree sequence  of ${\cal
  G}_u$  and  its  component  structure  by  a  randomization  process
denominated  {\em   local-swap}  \cite{Hanhijarvi09}.   Local-swap  is
performed  as follows:  We randomly  select an  existing  edge $\{v_i,
v_j\}$  of ${\cal  G}_u$ such  that the  two additional  edges $\{v_i,
v_k\}$ and  $\{v_j, v_l\}$ also  exist in ${\cal G}_u$,  provided that
$v_k,  v_i, v_j,  v_l$ are all different.  Then  we proceed  to  make the
rewiring, by generating the edges  $\{ v_i, v_l\}$ and $\{v_j, v_k\}$;
and removing the edges $\{v_i, v_k\}$ and $\{v_j, v_l\}$.  If $\{ v_i,
v_l\}$ or  $\{v_j, v_k\}$ already exist  in ${\cal G}_u$  we abort the
operation  and we randomly  select another  edge satisfying  the above
described  conditions  to  perform   the  local  swap.   According  to
\cite{Hanhijarvi09},  local-swap method can  perform all  rewirings of
links except those that imply the breaking of the component structure.
This  process  is  iteratively  repeated until  achieving  a  suitable
randomization  of  ${\cal  G}_u$  or after  a  predefined  number  of
iterations.
\item
Once the  local-swap randomization of  ${\cal G}_u$ is done,  we label
every node  with an arbitrary natural  number, from $1$  to $N$, being
$N$ the size of the graph. No repetitions are allowed.
\item
We now proceed  to define the arrows taking  into account the 
numbering  of the nodes defined in the previous step.   For every  pair of  connected nodes  in the
randomized version of ${\cal G}_u$,  we define the arrow from lower to
higher  number's  nodes. Formally,  given  a  undirected pair  $\{v_i,
v_j\}$ where  $i, j$  are the respective  labels obtained  through the
random   numbering,  if   $i>j$   then  $v_i\to   v_j$,  otherwise   $
v_i\leftarrow  v_i$. The  total order  of natural  numbers  avoids the
presence of cycles.
\end{enumerate}

Method  ${\mathbf {b}}$ consists of preserving the undirected degree sequence but not
preserving component structure. Component structure  is ensured by  step 2) in method   ${\mathbf {a}}$. In this  case, step 2) is replaced
by a direct rewiring process: $i)$ selecting a pair of different edges
$\{v_i,  v_k\},\{v_j,  v_l\}$ of  ${\cal  G}_u$;  $ii)$ Generate  with
probability $p=1/2$  either the  edges $\{v_i, v_l\},\{v_j,  v_k\}$ or
the edges  $\{v_i, v_j\}, \{v_l,v_k\}$  (provided that both  two edges
are  not  already  present)   and  $iii)$  remove  the  edges  $\{v_i,
v_k\},\{v_j, v_l\}$ -see fig. (\ref{random_method}b).

\subsection{Generating the ensemble from the directed degree sequence}

Beyond the randomization of the  raw topological structure of the real
DAG  conserving component structure,  one could  be interested  in the
preservation     of    the     directed    degree     sequence    -see
eq. (\ref{degreeSeq}).  This has an important physical interpretation,
since it implies that every node has an  invariant number of inputs
and  outputs, as  it  happens  with the  components  of an  electronic
device.  Under such  a  restriction we  can  no longer  work with  the
underlying graph but with the directed graph.

The proposed  algorithm, denoted  by method ${\mathbf {c}}$ -see fig.  (\ref{random_method}c)-, begins  with a
numbering  of the  nodes  resulting  from the  application  of a  {\em
  leaf-removal  algorithm}  \cite{Rodriguez-Caso2009}  and a  rewiring
operation constrained by  this numbering. Let us briefly  revise how a
leaf-removal algorithm works: From  the original graph, ${\cal G}$, we
iteratively  remove  the nodes  with  $k_{o}=0$  until the  complete
pruning  of  the   graph.   According  to  this,  a   DAG  can  be
\textit{layered}, and thus a partial order between nodes can be easily
established.   Formally,  the  $i$-th  iteration of  the  leaf-removal
algorithm  defines  the set  $V_i\subseteq  V$  of  nodes where  $V_i$
corresponds the  $i$-th {\em layer} of  the DAG. Then, any  DAG can be
redefined in terms of the resulting -ordered- layers of a leaf-removal
algorithm, i.e.,
\begin{equation}
W({\cal G})=\{W_1,...,W_l\}
\end{equation}
where no link between nodes of the same layer is established.
 
Method ${\mathbf{c}}$ -see fig. (\ref{random_method}c)- is defined as follows:
\begin{enumerate}
\item
Generate the set $W({\cal G})$ by applying the leaf removal algorithm.
\item
Perform a  random numbering  of the  nodes in such  a way  that, given
$v_i\in W_u$, and $v_k\in W_s$,
\begin{equation}
(u>s)\rightarrow(i>k) 
\end{equation}
\item
Select at random an edge $\langle v_k, v_j\rangle \in E$. Then we look
for the presence of two nodes  $v_i$, $v_l \in V$ by which either:
\begin{eqnarray}
&i)& \langle v_i, v_k\rangle ,\langle v_l, v_j\rangle \in E, \;\;{\rm or}\label{conf1}\\
&ii)& \langle v_k, v_i\rangle , \langle v_j, v_l\rangle \in E\label{conf2}.
\end{eqnarray}
Notice  that the  absence of  cycles makes these two  options mutually
exclusive.
\item
If the  condition (\ref{conf1}) is satisfied, the  pairs $\langle v_l,
v_k\rangle$ and  $\langle v_i, v_j\rangle$ are  generated and $\langle
v_i, v_k\rangle$,$\langle v_l,  v_j\rangle$ deleted, provided that the
following conditions  are satisfied: 1)$\langle  v_l, v_k\rangle\langle
v_i, v_j\rangle\notin  E$ and 2)$l>k$ and  $i>j$. If one  of these two
conditions does  not hold, the  rewiring event is aborted  and another
edge is newly selected at random.

\noindent 
If  condition  (\ref{conf2}) is  satisfied,  the  pairs $\langle  v_k,
v_l\rangle$  and  $\langle v_j,  v_i\rangle$  are generated,  deleting
$\langle v_k, v_i$, $\rangle  \langle v_j, v_l\rangle$ links, provided
that  $\langle  v_k, v_l\rangle\langle  v_j,  v_i\rangle\notin E$  and
$k>l$ and $j>i$  conditions are satisfied. Again, if  one of these two
conditions does not hold the rewiring event is restarted.
\end{enumerate}

Finally, the randomization method ${\mathbf { d}}$ preserves the
directed   degree  sequence   but  do   not  preserve   the  component
strucutre.  In  this  case,  step  3) is  replaced  by  the  following
procedure: $i)$  select two edges at random  $\langle v_k, v_i\rangle,
\langle v_l, v_j\rangle\in E$;  $ii)$ generate the edges $\langle v_k,
v_j\rangle,  \langle  v_l,  v_i\rangle$  provided that  $\langle  v_k,
v_j\rangle, \langle  v_l, v_i\rangle\notin E$ and  that $k>j,l>i$.  If
some of  these conditions does not  hold, process is aborted and
we restart  the  rewiring event.   $iii)$  If  conditions are  satisfied, 
 $\langle  v_k, v_i\rangle, \langle  v_l, v_j\rangle$ are removed
-see fig. (\ref{random_method}d).

\section{Exploring the randomness of DAGs}

In this  section we  apply the above  defined algorithms to  some real
topologies to construct an ensemble of randomized networks  (also known as  \textit{surrogate data} in other
scientific  communities) 
preserving the defined  topological  invariants.  First of all,  we need  to define  proper  measures to
evaluate the  level of randomness of  our systems.

\subsection{Testing the success of the randomization process}

As  it  is  described   above,  randomizations  are  subject  to  very
restrictive constraints  since not all  (un)-directed degree sequence configurations
are graphical.  Therefore, the  success of DAG randomization processes
must  be properly evaluated.   Two estimators  were measured  for this
purpose. First,  a \textit{dissimilarity} parameter  ${\cal D}$ is  proposed to
measure  how the  graph evolves  along the  iterations with respect  to the
original  one.  Second,  the  deterioration of  present  degree-degree
relations  is also  reported by  means of  an estimator  borrowed from
information    theory,    the    so-called   {\em    joint    entropy}
\cite{Shannon1948}.

\subsubsection{Dissimilarity}

The  dissimilarity parameter  ${\cal  D}$ between  two  graphs is  the
relative  frequency of link  mismatches between  them, i.e.,  the {\em
  Hamming distance} of their adjacency  matrices.  In the context of a
randomization process,  let us define  ${\cal A}$ and ${\cal  A}^t$ as
the adjacency matrices of an original graph (${\cal G}$) and the graph
resulting from the  application of $t$ randomization iterations (${\cal
  G}^t$), respectively.   Their dissimilarity can be expressed as
\begin{equation}
{\cal D}({\cal G},{\cal G}^t) \equiv \frac{1}{2|E|}\sum_{i,j} 1 - \delta({\cal A}_{ij},{\cal A}_{ij}^t),
\end{equation}
where   $\delta$ is the Kronecker's delta and $|E|$ denotes the number of  links of both ${\cal G}$ and ${\cal
  G}^t$, since the undirected degree  sequence is preserved in the four
randomization methods.

\subsubsection{Degree-degree joint entropy}

Given two random variables, $X,Y$, the {\em joint entropy} between $X$
and $Y$, $H(X,Y)$ is given by:
\begin{equation}
H(X,Y)=-\sum_{x,y}\mathbb{P}(x,y)\log(\mathbb{P}(x,y),
\label{U(X,Y)}
\end{equation}
being $\mathbb{P}(X,Y)$ the joint  probability of the pair of outcomes
$x,y$ happening together -throughout this paper $\log_2$ will be used. Let
us detail  how every concept is  translated in a useful  way to become
graph measures.

Joint  entropy for the  evaluation of  degree-degree relations  can be
expressed as
\begin{equation}
H({\cal G}_{u})=\sum_{i \geq j}\mathbb{P}(i,j)\log\mathbb{P}(i,j)
\label{undirected}
\end{equation}
where $\mathbb{P}(i,j)$ defines the  probability of finding a randomly
selected  link that  connects  two  nodes $v_m,  v_l\in  V$ such  that
$k(v_m)=i,\;k(v_l)=j$. This measurement was found to be more appropriate than other existing altervatives for the purpose of monitoring the degree-degree interplay along the randomization processes \footnote{A measure quantifiying  {\em how random or how deterministic is a structure in
relation to the space allowed by the topological invariants} was required. The
degree-degree joint entropy of a graph holds this property. Other valuable measures, 
such as assortativity \cite{Newman2002,Foster2009} or mutual information \cite{sole2004} have been pointed out.
Assortativity measures degree-degree correlations and degree-degree
mutual information quantifies the predictability of neghbors' degrees
from the sole knowledge of the degree of a given node in relation to
the {\em available} degree richness of the system. The former case
strictly looks for linear relationships and it is supposed to be a more
appropriate measure for normally distributed data. Furthermore both
approaches naturally require of certain degree-degree variance within
the graph. For instance, a  large feed-forward single chain of nodes
has a strong degree-degree determinism that none of these two
measurements would capture. The reason is that most of the
degree-degree pairs would be $(2,2)$ for undirected and $(1,1)$ for
any directed degree analyses. In this sense, degree-degree joint entropy
provides a suitable measure of the relation or determinism of
degree-degree relations with neither parametric assumptions nor
degree-degree variance requisites. Accordingly, we used the concept of degree-degree relations instead of
degree-degree correlations.\label{footmethod}}.  The subscript  "$u$" emphasizes  that  such a
measure does not take into account the directed nature of the graph. Joint  entropy  
quantifications   for  degree-degree  considering  the
directed degree sequence can  be easily  derived.  In this  case three
additional joint entropies  attending  directedness can be considered, namely the ones
accounting  for $k_{i}k_{o}$,  $k_{i}k_{i}$  and $k_{o}k_{o}$
relations.  Although  more elaborated definitions  of this probability
can be  proposed, for the sake  of simplicity we  assessed whether two
nodes  with given  degrees tend  to  be connected,  not matter  the
direction of  the arrow  connecting them. 
Then  the $(k_{i}k_{i})$-joint entropy  of a directed  graph ${\cal G}$,  $H^{i,i}({\cal G})$
is expressed as:
\begin{equation}
H^{i,i}({\cal G})=\sum_{k\geq j} \mathbb{P}_{i,i}(k,j)\log \mathbb{P}_{i,i}(k,j),
\label{Iii}
\end{equation}  
where $\mathbb{P}_{i,i}(k,j)$ is the probability of that a link chosen
at random connects a node with $k_{i}=i$ to another with $k_{i}=j$.
A similar expression is obtained for $H^{o,o}({\cal G})$. Finally, $H^{i,o}({\cal G})$
is defined as:
\begin{equation}
H^{i,o}({\cal G})=\sum_{k,j} \mathbb{P}_{i,o}(k,j)\log \mathbb{P}_{i,o}(k,j).
\label{Iio}
\end{equation}  
Notice     that    this is the only case where   $\mathbb{P}_{i,o}(k,j)\neq
\mathbb{P}_{i,o}(j,k)$.

The ensemble of  random graphs produced  from  a original  graph ${\cal  G}$
after $t$  iterations can   be  associated  to  the  undirected
degree-degree  joint entropy  distribution of  its  conforming graphs,
which   can   be   characterized   by  its   mean   $\langle   H({\cal
  G}_u^t)\rangle$   and   its   standard   deviation   $\sigma(H({\cal
  G}_u^t))$. The closeness of the  joint entropy value of the original
graph to the ensemble  distribution can be quantified by means
of the $Z$-score, which reads:
\begin{equation}
Z({\cal G}_u)=\frac{H({\cal G}_u)-\langle H({\cal G}_u^t) \rangle}{\sigma(H({\cal G}_u^t))}.
\end{equation}
 The statistical significance level was set at $p<0.001$ which
for  two  tails corresponds  to  $|Z|>3.27$.  Significant values  were
denoted  by $Z^*$  in the  tables describing  joint entropy  values of
graphs. Values  of $Z<3.27$ means that the  degree-degree relations at
the original network ${\cal G}$  are significantly high respect to the
$Z-$distribution of its random ensemble. Values of $Z>3.27$ means that
the  degree-degree relations at  the original  network ${\cal  G}$ are
significantly  low  respect  to  the $Z-$distribution  of  its  random
ensemble.  Finally, values  within the  range  $[-3.27,3.27]$ indicate
that no  significant differences  in the degree-degree  relations were
found    between    the   original    graph    and   its    randomized
ensemble.   Analogously,  we   can   compute  $\langle   H^{i,o}({\cal
  G}^t)\rangle$,   $\langle   H^{i,i}({\cal  G}^t)\rangle$,   $\langle
H^{o,o}({\cal G}^t)\rangle$ and its  associated $Z$-scores at the step
$t$ of the randomization process.

\begingroup 
\begin{figure*}
\begin{center}								
\includegraphics[width=13cm]{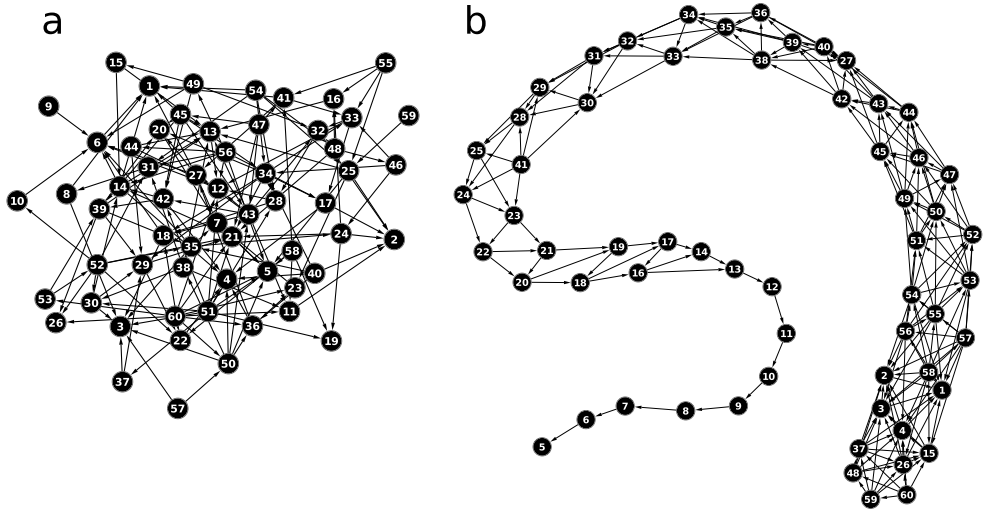}		
 \caption{Illustration of a \textit{Random}-DAG  with $N=60$  and  $\langle  k \rangle  =3$
   (a). Illustration of a \textit{Snake}-DAG  with  $N=60$  and $K=6$ (b). In both graphs arrows go from nodes with larger numbers to nodes with smaller number.}
\label{snake}	
\end{center}								
\end{figure*}
\squeezetable
\begin{table*}
\caption{\label{tab:table_random_dag}
Joint entropy values for a \textit{Random}-DAG  of $|V|=600$
  and $|E|=6000$  and a  set of $500$  randomized graphs  after $2^{18}$
  iterations of  each of  the four randomization  methods (alphabetically
  denoted). Symbol ($^*$) denotes significant differences.}
\begin{ruledtabular}
\begin{tabular}{ccllll}
\textrm{method}&
\textrm{${\cal  D}$}&
\textrm{$H^u({\cal G}_u^t)$}&
\textrm{$H^{i,o}({\cal G}^t)$}&
\textrm{$H^{i,i}({\cal G}^t)$}&
\textrm{$H^{o,o}({\cal G}^t)$}\\
\colrule
${\cal G}$ orig.       & -     & 8.279                       & 9.214                         & 9.163                         & 9.250                     \\
${\mathbf { a}}$ & 0.98  & 8.283 $\pm$ 0.003 (Z=-1.67) & 9.21 $\pm$ 0.02 (Z=0.25)       & 9.20 $\pm$ 0.04 (Z=-0.95)       & 9.21 $\pm$ 0.04   (Z=1.01) \\
${\mathbf { b}}$ & 0.98  & 8.283 $\pm$ 0.003 (Z=-1.46) & 9.21 $\pm$ 0.03 (Z=0.21)       & 9.20 $\pm$ 0.04 (Z=-0.90)       & 9.21 $\pm$ 0.04   (Z=1.02) \\
${\mathbf { c}}$ & 0.96  & 8.282 $\pm$ 0.003 (Z=-1.03) & 9.209 $\pm$ 0.003 (Z=1.92)     & 9.157 $\pm$ 0.004 (Z=1.76)     & 9.249 $\pm$ 0.004 (Z=0.03) \\
${\mathbf { d}}$ & 0.96  & 8.281 $\pm$ 0.003 (Z=-0.62) & 9.206 $\pm$ 0.002 (Z=2.96)     & 9.152 $\pm$ 0.004 (Z=3.03)     & 9.247 $\pm$ 0.004 (Z=0.68) \\
\end{tabular}
\end{ruledtabular}
\end{table*}
\squeezetable
\begin{table*}
\caption{%
Joint entropy values for a \textit{snake}-DAG ($|V|=600$ and $K=6$) and a set of $500$ randomized graphs after $2^{18}$ iterations of each of the four randomization methods (alphabetically denoted). Symbol ($^*$) denotes significant differences.
}
\begin{ruledtabular}
\begin{tabular}{ccllll}
\textrm{method}&
\textrm{${\cal  D}$}&
\textrm{$H^u({\cal G}_u^t)$}&
\textrm{$H^{i,o}({\cal G}^t)$}&
\textrm{$H^{i,i}({\cal G}^t)$}&
\textrm{$H^{o,o}({\cal G}^t)$}\\
\colrule
${\cal G}$ orig.       & -    & 2.998                               & 4.331                              & 4.131                              & 4.161                              \\
${\mathbf { a}}$ & 0.99 & 5.281  $\pm$ 0.003  (Z=-691.88$^*$) & 6.98   $\pm$ 0.02   (Z=-152.2$^*$) & 6.97  $\pm$ 0.03   (Z=-91.53$^*$) & 6.97   $\pm$ 0.03   (Z=-86.86$^*$) \\
${\mathbf { b}}$ & 0.99 & 5.281  $\pm$ 0.003  (Z=-671.59$^*$) & 6.98   $\pm$ 0.02   (Z=-174.3$^*$) & 6.97  $\pm$ 0.03   (Z=-91.83$^*$) & 6.97   $\pm$ 0.03   (Z=-92.05$^*$) \\
${\mathbf { c}}$ & 0.95 & 4.96 $\pm$ 0.03 (Z=-58.88$^*$)  & 5.37 $\pm$ 0.02 (Z=-52.96$^*$) & 5.50 $\pm$ 0.02 (Z=-63.42$^*$) & 5.12  $\pm$ 0.02 (Z=-42.22$^*$) \\
${\mathbf { d}}$ & 0.93 & 4.71  $\pm$ 0.03 (Z=-51.63$^*$)  & 5.15 $\pm$ 0.02 (Z=-43.89$^*$) & 5.22 $\pm$ 0.03  (Z=-41.95$^*$) & 5.05 $\pm$ 0.02 (Z=-40.23$^*$) \\
\end{tabular}
\end{ruledtabular}
\label{table_snake}
\end{table*}
\endgroup

\subsection{Extreme Graphs}

Prior  to evaluate  the  randomness  of real  DAGs,  we construct  two
extreme topologies in order to evaluate the behavior of the algorithms
using the measures defined  above. The first model, random-DAG, permit
us  to   test  the  randomization  methods  in   a  highly  disordered
degree-degree  scenario.  In  terms  of degree-degree  joint  entropy,
minimal changes  are expected along the  randomization processes.  The
second  model, \textit{snake}-DAG,  permit us  the same  test but  in  a highly
ordered scenario where large increments of joint entropy values should
be observed.

\subsubsection{\textit{random}-DAG model}

The  first one  is a  completely degree-degree  disordered DAG,  up to
finite  size effects. Let  $V$ be  a set  of $N$  nodes, by  which the
probability for two  of them to be connected is  constant and equal to
$p$. This  is the definition of the Erd\"os  R\'enyi (ER) graph. Once we  have an ER
graph, we  randomly label the nodes  of $V$ sequentially,  from $1$ to
$N$. Finally, we define the direction  of the arrows by looking at the
labeling  of the nodes  and observing  condition (\ref{i>j}).  We will
refer to this model as \textit{random}-DAG.

We  created a  random-DAG of  $|V|=600$ and  $|E|=6000$ and,  for each
method,  an ensemble of  $500$ randomized  graphs product  of $2^{18}$
iterations. See fig. (\ref{snake}a) for an example of this graph. As shown  in  Table \ref{tab:table_random_dag},  none of  the
degree-degree relations of  the random-DAG where neither significantly
low  nor significantly  high with  respect  to any  of its  randomized
ensembles. This  result indicates  that the four  randomization methods
proposed  here do not  produce significant  undesirable biases  in the
degree-degree almost null relations of a originally random-DAG.

\subsubsection{\textit{snake}-DAG model}

Opposed to  the random-DAG model, we construct  a highly degree-degree
ordered  acyclic  graph,  which  we  will  call  {\em  snake}-DAG  see
fig. (\ref{snake}). In this graph, nodes of the same degree tend to be
connected among them, giving rise to a high degree-degree relation and
thus very low joint entropy values.  In the following lines we outline
the construction of this network.

Let us  consider $K\equiv k_{o}^{max}$  as the highest  outdegree to
appear in the  resulting graph. Let $V$ be the set  of nodes such that
there exists an integer $n$ by which $n\cdot K=|V|$. We then perform a
partition of $V$ in $K$ different subsets
\begin{equation}
{\cal P}(V)=\{V_1,...,V_K\}. 
\end{equation}
In  this  partition,  for  any  $V_i\in {\cal  P}(V)$,  $|V_i|=n$.  We
sequentially number the nodes of the set $V$ is the following way: For
the subset  of nodes $V_1$, the label  will run from $1$  to $n$, thus
obtaining:
\begin{equation}
V_1=\{v_1,...,v_n\}\nonumber
\end{equation}
For the  subset of  nodes $V_{2}$,  the label will  run from  $n+1$ to
$2n$:
\begin{equation}
V_2=\{v_{n+1},...,v_{2n}\}.\nonumber
\end{equation}
We follow the numbering by using the criteria that the nodes of subset
$V_i$ will be  labeled from $(i-1)n+1$ to $i\cdot  n$, until
all the nodes of $V$ are numbered. We then identify the label of the partition with the out-degree of the
nodes belonging to it, namely:
\begin{equation}
(v_i\in V_m)\Rightarrow (k_{o}(v_i)=m).
\label{kout=m}
\end{equation} 
Now we  proceed to  define the connections:  For any $v_i\in  V_m$, we
will have the  following links $\langle v_i,v_{i-1}\rangle,...,\langle
v_i,v_{i-m}\rangle$.  This process  excludes node  $v_{1}$  which will
only  receive a link  from $v_2$.  We observe  that, in  general, both
$v_1$ and $v_2$ belong to $V_1$. Finally,  to  break the  extreme  symmetry  of  the obtained  net,  we
introduce a  minimal source of  noise by renumbering a  small fraction
$\sim 0.05$ of the nodes with a further arrow orientation consistently
with the new numbering, as depicted in eq.  (\ref{i>j}).

Analogous to the experiment performed  with a \textit{random}-DAG, we create a
\textit{snake}-DAG  of  $|V|=600$  and  $|E|=2099$ ($K=6$) and, for  each  method,  an
ensemble of $500$ randomized graphs product of $2^{18}$ iterations. See fig. (\ref{snake}b) for an example of this graph. As
shown in  Table \ref{table_snake}, all the  degree-degree relations of
the  \textit{snake}-DAG were  significantly high  with respect  to any  of the
randomized ensembles. This result indicates that the four randomization
methods proposed  here are able  to successfully deteriorate  the high
degree-degree relations present at the \textit{snake}-DAG.

\subsection{Real biological and social DAGs}

Results  of previous  section have  checked  the behavior  of the  four
methods   in  two  toy models  with  high   and  low
degree-degree relations  respectively. In  this section we  proceed to
evaluate three DAGs representing real systems: the \textit{C. elegans}
cell  lineage  network,  the  Milgram's  citation network  and  a  PhD
student-advisor network.
\subsubsection{\textit{C. elegans} cell  lineage network} 
\begingroup
\begin{figure*}
\begin{center}								
\includegraphics[width=18cm]{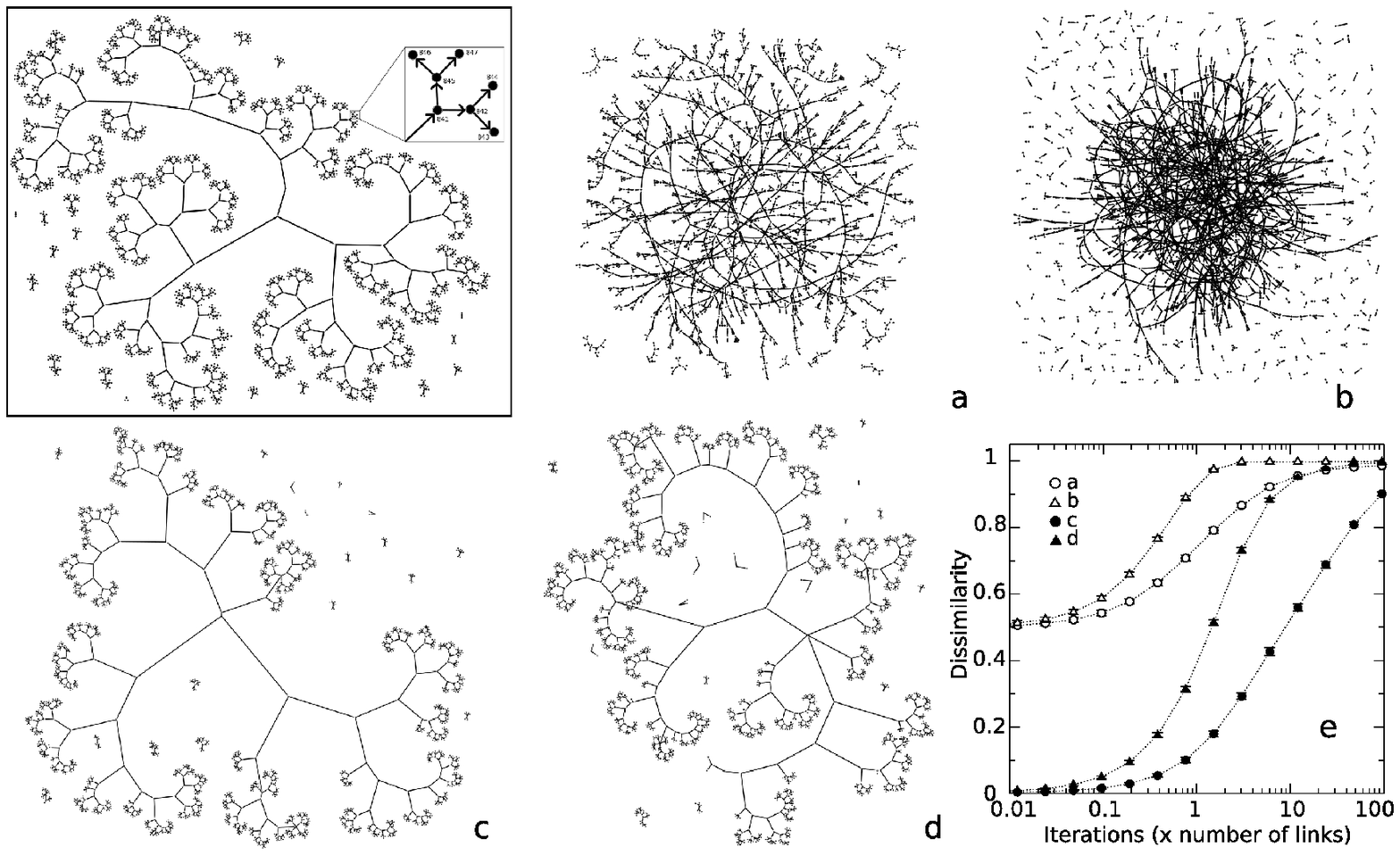}		
 \caption{DAG    representation    of     the    cell    lineage    of
   \textit{Caebnorhabditis  elegans} (inbox). A  prototypic  randomized  network  after
  $2^{18}$ iterations is present for
   each method: randomization   preserving  undirected   degree
     sequence and component  size distribution (a), randomization only
     preserving  the  undirected  degree sequence  (b),  randomization
     preserving   directed   degree   sequence  and   component   size
     distribution  (c)  and  randomization  preserving  only  directed
     degree  sequence  (d).   Panel  (e)  represents  the  dissimilarity
     preserving   the   original   network   along  the   process   of
     randomization for  every randomization  type. The mean  and the
   standard deviation  of $500$ graph randomizations  are shown
   for each point.}
\label{c_elegans}	
\end{center}								
\end{figure*}
\squeezetable
\begin{table*}
\caption{\label{tab:table_celegans}%
Joint entropy values for the original \textit{C. elegans} network and a set of $500$ randomized networks after $2^{18}$ iterations of each of the four randomization methods (alphabetically denoted). Symbol ($^*$) denotes significant differences and ($^a$) denotes that Z-score is not computable due to $\sigma=0$
}
\begin{ruledtabular}
\begin{tabular}{ccllll}
\textrm{method}&
\textrm{${\cal  D}$}&
\textrm{$H^u({\cal G}_u^t)$}&
\textrm{$H^{i,o}({\cal G}^t)$}&
\textrm{$H^{i,i}({\cal G}^t)$}&
\textrm{$H^{o,o}({\cal G}^t)$}\\
\colrule
${\cal G}$ orig.       & -    & 1.832                            & 0.991                           & 0.116                            & 1.732\\
${\mathbf { a}}$ & 0.99 & 1.833 $\pm$ 0.001 (Z=-0.58)      & 3.71 $\pm$ 0.01   (Z=-205.9$^*$) & 3.67   $\pm$ 0.02 (Z=-167.4$^*$) & 3.66  $\pm$ 0.02 (Z=-92.88$^*$ )\\
${\mathbf { b}}$ & 1.00 & 1.961 $\pm$ 0.001 (Z=-117.3$^*$) & 3.71 $\pm$ 0.01   (Z=-224.5$^*$) & 3.69   $\pm$ 0.02 (Z=-175.0$^*$) & 3.68  $\pm$ 0.02 (Z=-92.55$^*$) \\
${\mathbf { c}}$ & 0.90 & 1.831 $\pm$ 0.002 (Z=0.41)       & 0.990 $\pm$ 0.001 (Z=1.33)       & 0.116  $\pm$ 0.0$^a$         & 1.736 $\pm$ 0.002 (Z=-2.67)\\
${\mathbf { d}}$ & 0.99 & 1.832 $\pm$ 0.002 (Z=0.80)      & 0.989 $\pm$ 0.001 (Z=-1.6)       & 0.116 $\pm$ 0.0$^a$        & 1.734 $\pm$ 0.002 (Z=-1.67)\\
\end{tabular}
\end{ruledtabular}
\end{table*}
\endgroup 

The  first system  chosen  is  a cell  lineage  network.  Briefly,  it
captures  the genealogic  pedigree  of cells  related through  mitotic
division during  its development in  a tree-like structure.   The cell
lineage  network  of  \textit{C.   elegans}  was  retrieved  from  the
WormBase~\footnote{http://www.wormbase.org, release WS202, date June 03
  2009} \textit{C.  elegans} repository.   In this network the initial
egg  division (the  giant component)  and  alternative variants  of
neural post-embrionic cell lines  are included in a 18-component graph
representation.   All the  randomization  methods were  applied up  to
$2^{18}$ iterations. The dissimilarity values reached were over $0.90$
in all cases, indicating a  successful alteration of most of the links
under the different topological invariants.

Figure  (\ref{c_elegans}) shows  the original  and  a representative
DAG  for  each randomization  method.   Note  the deep  fragmentation
produced by  method ${\mathbf {b}}$, where  only the undirected
degree sequence is preserved ($205$ components in fig. (\ref{c_elegans}b) against 18-component in  the original graph).  Interestingly, figure (\ref{c_elegans} d) shows that the tree-like structure and the number of
graph components  are conserved by  just only preserving  the directed
degree sequence invariant.  The reason  is that the regular pattern of
$k_{i}=1$ for  all non-maximal nodes in its  directed degree sequence
is graphical  only in a  tree structure. In this  particular situation
the  number of  DAG components  coincides with  the number  of maximal
nodes. However, the  size of components is not  strictly conserved and
this condition is only possible by applying the local-swap as observed
in figures \ref{c_elegans}a and  \ref{c_elegans}c.

Looking at figure (\ref{c_elegans}e),  all randomizations provide
completely re-allocation  of nodes. It  is worth to note  that methods
not preserving the directed degree sequence start with a dissimilarity
values of  around $0.5$.  The reason  is that a  complete random arrow
orientation occurs  for every  iteration.  This contrasts  with methods
${\mathbf { c}}$ and ${\mathbf {d}}$ where rewiring operated  over the directed
graph.   Table   (\ref{tab:table_celegans})  shows  that   almost  all the
ensembles generated  through the proposed  algorithms display relevant
deviations in the values of joint entropies. They are higher than the
observed  in the  real DAG when the  undirected degree sequence  is preserved   (methods   $\mathbf{{a},{b}}$).
Otherwise, when the directed
degree  sequence is conserved,  no deviations  are  found
(methods $\mathbf{{c},{d}}$) due to the very restrictive (even zero)  standard deviations.

\subsubsection{Milgram's citation network}

The second system is a sample  of the process of article citation. The
chosen system used to illustrate this process is the resulting network
containing  the papers  that cite  "S Milgram's  1967  \textit{Psychology Today}"
paper  or  use \textit{Small  World } in  the title.  This  network  was
retrieved  from  to Pajek's  network  dataset \footnote{V. Batagelj  and
A. Mrvar      (2006). Pajek    datasets.
  http://vlado.fmf.uni-lj.si/pub/networks/data/)}.        All      the
randomization  methods were  applied up  to $2^{18}$  iterations (see figure \ref{citations}). The
dissimilarity values  reached were  over $0.86$ for  methods ${\mathbf
  { a}}$ and ${\mathbf {  b}}$ and over $0.76$ for methods
${\mathbf  { c}}$  and ${\mathbf  { d}}$.   This indicates
that keeping  the directed degree sequence as  a topological invariant
reduces   the  heterogeneity   of   feasible  graphs   and  thus   the
dissimilarity reached.

Table   (\ref{tab:table_citations})    shows    significantly   low
joint-entropy   values,   indicating   that   this  DAG   displays   a
statistically relevant undirected degree-degree relation respect to
all their  randomized ensembles.  It can also  be appreciated relevant
indegree-outdegree  and outdegree-outdegree relations  with respect
to all  their randomized ensembles.  In the  case of indegree-indegree
entropies,  it shows  a  high degree-degree relation  respect  to the  randomized
ensembles  produced by  methods ${\mathbf  { a}}$  and ${\mathbf
  {  b}}$,  while  there  were  no differences  respect  to  the
randomized ensembles  produced by  methods ${\mathbf {  c}}$ and
${\mathbf { d}}$.   This result indicates that indegree-indegree
relations in  the Milgram's citation  network are high respect  to the
random DAGs that preserve only  the undirected degree sequence and not
differentiable  from  the  ones  obtained by
preserving the directed  degree sequence.
\begingroup
\begin{figure*}
\begin{center}	
\includegraphics[width=18cm]{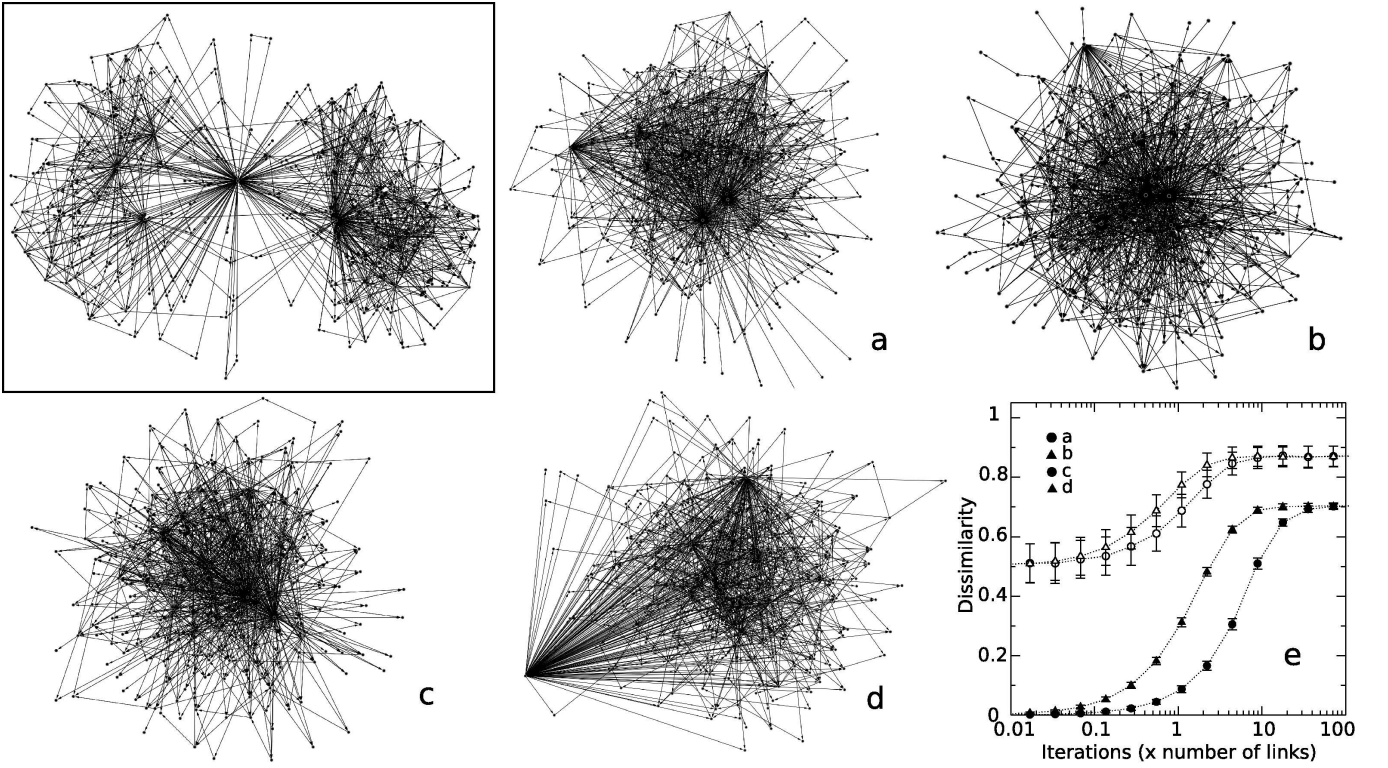}		
 \caption{DAG  representation   of  the  Milgram's   citation  network
   (inbox).  A  prototypic  randomized  network  after
  $2^{18}$ iterations is present for
   each method:  randomization preserving undirected
   degree sequence and  component size distribution (a), randomization
   only preserving  the undirected degree  sequence (b), randomization
   preserving directed degree sequence and component size distribution
   (c)  and  randomization preserving  only  directed degree  sequence
   (d). Panel  (e) represents the dissimilarity  respecting the original
   network along the process  of randomization for every randomization
   type.   The  mean  and   the  standard  deviation  of  $500$  graph
   randomizations are shown for each point.}
\label{citations}	
\end{center}								
\end{figure*}
\squeezetable
\begin{table*}
\caption{\label{tab:table_citations}
  Joint entropy values for the original \textit{Milgram's citation} network
 and a set of $500$ randomized networks after
  $2^{18}$  iterations  of  each  of  the  four  randomization  methods
  (alphabetically denoted). Symbol ($^*$) denotes significant differences. }
\begin{ruledtabular}
\begin{tabular}{ccllll}
\textrm{method}&
\textrm{${\cal  D}$}&
\textrm{$H^u({\cal G}_u^t)$}&
\textrm{$H^{i,o}({\cal G}^t)$}&
\textrm{$H^{i,i}({\cal G}^t)$}&
\textrm{$H^{o,o}({\cal G}^t)$}\\
\colrule
${\cal G}$ orig.       & -    & 9.03                         & 7.38                           & 7.16                           & 7.53 \\
${\mathbf { a}}$ & 0.87 & 9.24$ \pm$ 0.02 (Z=-10.5$^*$) & 8.2 $\pm$ 0.1  (Z=-8.6$^*$)  & 8.2 $\pm$ 0.1 (Z=-7.8$^*$)   & 8.20 $\pm$ 0.1  (Z=-4.8$^*$)\\
${\mathbf { b}}$ & 0.87 & 9.24 $\pm$ 0.02 (Z=-10.5$^*$) & 8.2 $\pm$ 0.1  (Z=-8.6$^*$)  & 8.2 $\pm$ 0.1 (Z=-7.6$^*$)   & 8.22 $\pm$ 0.1  (Z=-5.1$^*$)\\
${\mathbf { c}}$ & 0.70 & 9.18 $\pm$ 0.02 (Z=-7.35$^*$) & 7.45 $\pm$ 0.01 (Z=-7.38$^*$)  & 7.17 $\pm$ 0.01 (Z=-1.42)      & 7.63 $\pm$ 0.01 (Z=-8.59$^*$)\\
${\mathbf { d}}$ & 0.70 & 9.18 $\pm$ 0.02 (Z=-7.93$^*$) & 7.45 $\pm$ 0.01 (Z=-7.34$^*$)  & 7.17 $\pm$ 0.01 (Z=-1.40)      & 7.63 $\pm$ 0.01 (Z=-8.63$^*$)\\
\end{tabular}
\end{ruledtabular}
\end{table*}
\endgroup

This example  illustrates how  a randomization process  destroys local
associations and the heterogeneous  partition observed in the original
DAG (figure  \ref{citations} inbox).   In this case,  due to  the high
connectivity of  the original DAG,  -it is worth  to note that  such a
graph contains  several nodes  whose connectivity is  ${\cal O}(|V|)$-
fragmentation is unlike to happen  due to a high average connectivity.
Furthermore, as a  side effect, an upper boundary  below maximal value
of  dissimilarity is  imposed  depending on  the randomization  method
used.  This is due to a considerable fraction of failed rewiring attempts.  An  example of that is provided by a clique conformation where no rewiring is possible. In this case no  effective of rewiring can be done since all
possible  link combinations satisfying  the directed  acyclic condition
are actually in the network.

\subsubsection{PhD student-advisor network}

The last system evaluated in  this paper contains the ties between PhD
students and their advisors  in theoretical computer science. Each arc
points  from an  advisor to  a student.   Data was  retrieved  from to
Pajek's   network  dataset \footnote{V.  Batagelj   and   A. Mrvar
(2006). Pajek datasets.
  http://vlado.fmf.uni-lj.si/pub/networks/data/esna/CSPhD.html}.  This
network illustrates  just the  intermediate situation between  the two
previous  examples.  It  is  a DAG  able  to be  fragmented (when  DAG
components conservation  is not imposed) but  with \textit{just right}
connectivity:  too low  to avoid  fragmentation  but not  too high  to
impose an upper  bounding in dissimilarity, being, jointly  to the 
\textit{random}-DAG studied above, the DAG  structure closer to the assumptions of the
configuration  model.    Interestingly,  contrasting  with   the  {\em
  C. elegans} case, network  fragmentation also occurs when preserving
directed  degree  sequence but  not  when  the component  distribution
conservation is preserved.  All the randomization methods were applied
up to $2^{18}$ iterations. The dissimilarity values reached where over
$0.97$ in all cases, indicating a successful alteration of most of the
links under the different topological invariants.
\begingroup
\begin{figure*}
\begin{center}	
\includegraphics[width=18cm]{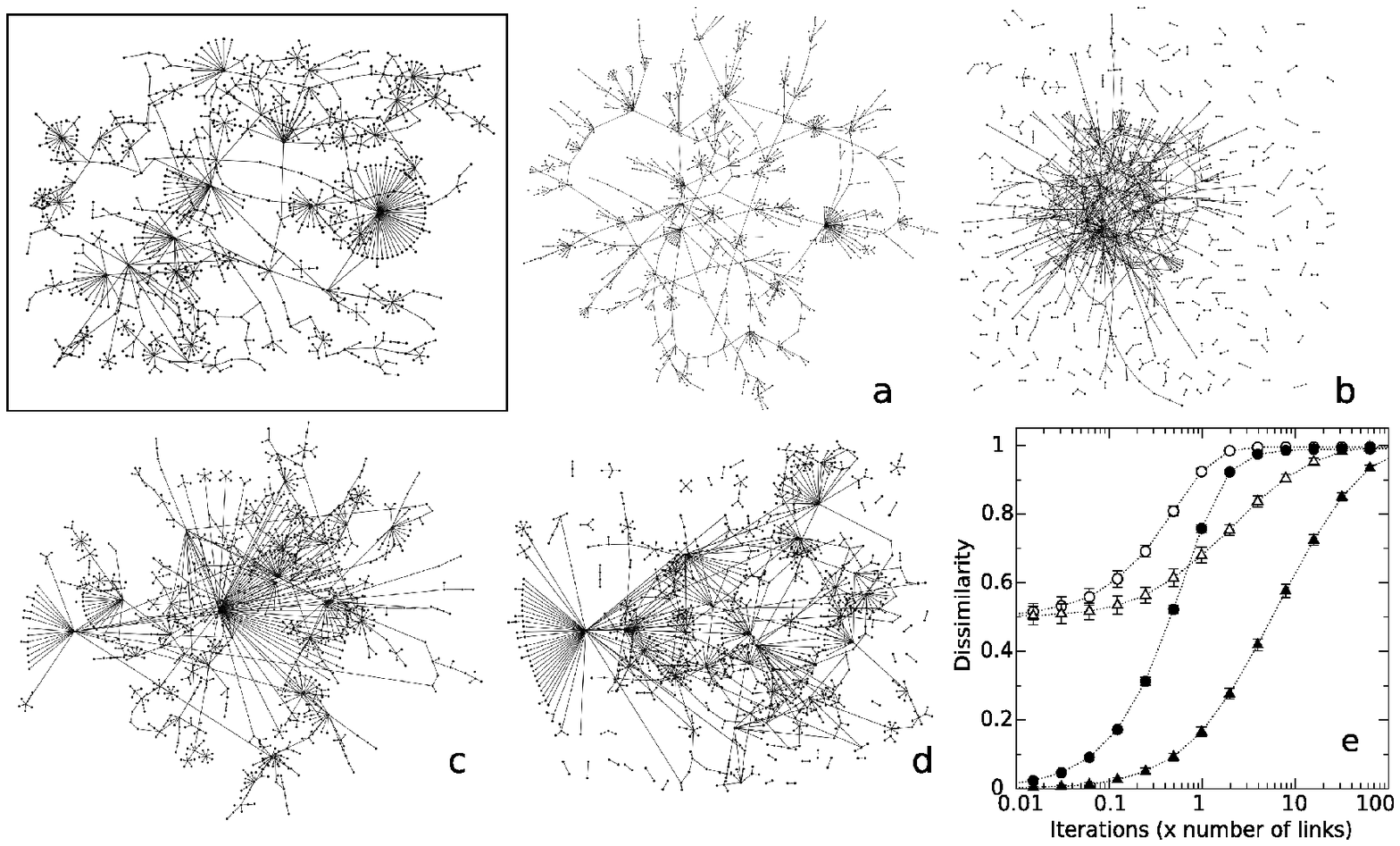}		
 \caption{DAG  representation of  PhD advisors  (inbox).  A  prototypic  randomized  network  after
  $2^{18}$ iterations is present for
   each method:
   randomization preserving  undirected degree sequence  and component
   size distribution (a), randomization only preserving the undirected
   degree  sequence  (b),  randomization  preserving  directed  degree
   sequence  and  component size  distribution  (c) and  randomization
   preserving only  directed degree  sequence (d). Panel  e represents
   the dissimilarity respecting the original network along the process
   of  randomization for every  randomization type.  The mean  and the
   standard deviation of $500$ graph randomizations are shown for each
   point.}
\label{real_data2}	
\end{center}								
\end{figure*}
\squeezetable
\begin{table*}
\caption{\label{tab:table_phd}
  Joint   entropy values  for the  \textit{PhD student-advisor}  network a set
  of $500$ randomized networks after  $2^{18}$ iterations of each of the
  four randomization methods (alphabetically denoted). Symbol ($^*$) denotes significant differences. }
\begin{ruledtabular}
\begin{tabular}{ccllll}
\textrm{method}&
\textrm{${\cal  D}$}&
\textrm{$H^u({\cal G}_u^t)$}&
\textrm{$H^{i,o}({\cal G}^t)$}&
\textrm{$H^{i,i}({\cal G}^t)$}&
\textrm{$H^{o,o}({\cal G}^t)$}\\
\colrule
${\cal G}$ orig.       & -    & 6.42                          & 4.075                            & 1.348                            & 6.34\\
${\mathbf { a}}$ & 0.99 & 6.47 $\pm$ 0.01 (Z=-4.50)      & 5.67  $\pm$ 0.06  (Z=-26.06$^*$) & 5.58  $\pm$ 0.1   (Z=-36.07$^*$) & 5.56 $\pm$ 0.1  (Z=6.70$^*$)\\
${\mathbf { b}}$ & 0.99 & 6.75 $\pm$ 0.01 (Z=-29.62$^*$) & 5.69  $\pm$ 0.06  (Z=-28.68$^*$) & 5.63  $\pm$ 0.1   (Z=-43.18$^*$) & 5.62 $\pm$ 0.1  (Z=7.46$^*$)\\
${\mathbf { c}}$ & 0.97 & 6.47 $\pm$ 0.01 (Z=-5.34)      & 4.076 $\pm$ 0.005 (Z=-0.32)      & 1.31  $\pm$ 0.01  (Z=3.33$^*$)   & 6.39 $\pm$ 0.01 (Z=-4.53$^*$)\\
${\mathbf { d}}$ & 0.98 & 6.59 $\pm$ 0.01 (Z=-13.55$^*$) & 4.072 $\pm$ 0.005 (Z=0.63)       & 1.366 $\pm$ 0.002 (Z=-8.27$^*$)  & 6.39 $\pm$ 0.01 (Z=-5.13$^*$)\\
\end{tabular}
\end{ruledtabular}
\end{table*}
\endgroup

Table   \ref{real_data2}   displays   statistically  significant   low
joint-entropy values, indicating that this DAG has relevant undirected
degree-degree  relations  with   respect  to  all  their  randomized
ensembles. It also displays a statistically relevant indegree-indegree
and  outdegree-outdegree  relations   with  respect  to  all  their
randomized  ensembles.  In  the case  of indegree-outdegree,  it shows
significant  relations  with respect  to  the randomized  ensembles
produced by methods ${\mathbf { a}}$ and ${\mathbf { b}}$, while there
were no  differences respect to  the randomized ensembles  produced by
methods ${\mathbf { c}}$ and  ${\mathbf { d}}$.  This result indicates
that  indegree-outdegree  relations   in  the  PhD  student-advisor
network are  high with respect  to the random  DAGs that preserve  only the
undirected degree  sequence and are  not differentiable from  the ones
obtained  by  random  DAGs  that  also preserve  the  directed  degree
sequence.

\section{Discussion} 

In this paper we present a set of four
algorithms based on an iterative process of rewiring for the construction of DAG random models. The difference between algorithms stems from two
topological invariants under consideration, namely, the conservation of the directed degree sequence and/or the conservation of the connected component distribution.
In contrast to other methods of random model construction, this approach works within the space of \textit{graphical}
solutions providing a feasible computational approximation for the exploration of such a graphical space considering a
defined number of topological invariants in the null-model ensemble generation.

Our methodology was evaluated through the analyses of both extreme and real graphs comparing them with their associated randomized ensembles using two  measures: dissimilarity and joint entropy.
While the former indicates whether connections are actually changed after 
randomization, the latter quantifies the
\textit{disorder} or uncertainty in the degree-degree relations, thereby being an 
indicator of randomness. In this context, it is worth to mention that other
measures such assortative mixing \cite{Newman2002,Foster2009} or mutual 
information \cite{sole2004} have been suggested for the evaluation of
degree-degree relations. In essence, these measures compare the actual degree correlations
 relation in the graph with the
expected one obtained from the \textit{remaining degree} information. 
A problem arises when a proper definition of remaining degree attending 
directedness needs of the information of the directed degree sequence 
because of the latter is a topological feature not preserved in all of our 
methods (methods ${\mathbf{a}}$ and ${\mathbf{b}}$).   Therefore, measures based on remaining
degree information, although extensively used as estimators of degree-degree relations in the network literature \cite{Newman2002,Foster2009,sole2004} cannot be applied 
in this work for a comparative evaluation of our methodology.

To overcome these limitations, joint entropy was used as a raw measure of uncertainty once defined to be applied to directed graphs
leading to four alternative descriptors according to in and out-degree information. Furthermore, the significance of
the variation of degree-degree relations between the random ensembles and the original graph was evaluated using a
Z-score estimator. The analysis of network models verified that our methods do not produce
a bias when applied to the \textit{random}-DAG model whilst they produced a significative increase of disorder of the
degree-degree relations on the \textit{snake}-DAG model when randomized -see table \ref{table_snake}. Going to
real systems, our analyses revealed that all the methods produced an $H^u$ greater than its respective original value, suggesting that randomizations disorder the underlying graph and they do not only affect the pattern of arrows. However, values of undirected joint entropies are different among methods, suggesting that  conservation of the DAG condition and the remaining topological invariants have a variable impact on the underlying network. When we look at the
\textit{directed} joint entropies a general increase of the values was observed for all of methods, although some exceptions
are observed. This is the case of PhD student-advisor network where the original network
exibits a diversity of degree-degree relations higher than the randomized ensemble.

Additionaly, our results show that preserving the component size structure is an important aspect to take 
into account since it has drammatic effects when the network
is markedly  sparse. This is the case of  \textit{C. elegans} and PhD student-advisor DAGs by which randomizations not preserving the component size produced a graph  fragmentation. On the contrary, high average degree guarantees the preservation of the giant component and randomization methods. In such circumstances, methods ${\mathbf{a}}$ and ${\mathbf{b}}$ give comparable values of joint entropies. Analogously, this is also sobserved for method ${\mathbf{c}}$ and ${\mathbf{d}}$  (see joint entropy for network models and also for the Milgram's citation
network). This feature indicates that the conservation of component structure premise is not relevant and produce indistinguisable topologies when graph fragmentation is unlike to happen.

Another important observation is related to the  small values displayed by standard deviations in joint entropies. For methods ${\mathbf{c}}$ and ${\mathbf{d}}$  are one order of magnitud
lower than the ones obtained for methods ${\mathbf{a}}$ and ${\mathbf{b}}$ . It suggests that just directed degree sequence conservation is enough to severely reduce the space of graphical configurations.
Consistenly, it was observed that, in general, methods ${\mathbf{c}}$ and ${\mathbf{d}}$ provided lower  Z-values than  ${\mathbf{a}}$ and ${\mathbf{b}}$. However, the small divergence of the obtained values
is not explained by a non-effective rewiring since high values of dissimilarity were reached. An interesting exception was found in the Milgram's citation network where dissimilarity values after
processes of randomization were markedly lower than the observed in the other real networks, as well as in the toy models.
An explanation can be found in the presence of {\em superhubs}, nodes whose connectivity is $\sim {\cal O}(|V|)$.
This introduces a strong constraint in the rewiring, difficult -even impossible- to overcome. Nevertheless figure
(\ref{citations}) illustrates that the original network seems to be partitioned in two regions. Using the same layout for randomized graphs, we observed that such a partition was lost, suggesting that rewiring process was accounted. Contrasting to this behaviour, \textit{C.  elegans} randomized ensembles were completely suffled
-as indicated by the high values of dissimilarity- but degree-degree relations were not always significatively altered. This
is specially evident when directed degree sequence is conserved. An explanation can be obtained by the fact that
this DAG is practically a dichotomic tree. This network is sparse enough to be fragmented as it happened in method ${\mathbf{b}}$. Nevertheless, preserving its extremal directed degree sequence was enough to conserve the number of components (not their size though). This is in aggreement to the constraint in the number of DAG components described in eq. (\ref{boundary}).
In fact, this is a result of the limited space of possibilities permited by the extremal directed degree sequence and therefore very little variation is found in the joint entropies (notice the case of zero
$H^{i,i}$ for method ${\mathbf{c}}$). Interestingly, when directed degree sequence and
component structure are not preserved, tree configuration is unlikely to happen by chance. However, tree structure is practically the only
solution when directed degree sequence is preserved even not conserving the component size distribution.

The  choice of
topological constraints (i.e. the particular method ) for a desired randomization process depends upon  the question the researcher wants to explore, rather than upon a technical issue.
Preserving the directed degree sequence captures the need to fix the
number of  inputs and outputs for every element. Randomizations attending to this constraint  (for example, in a technological system) may be interpreted as a rewiring of an
electronic circuit  by a random assembling  of integrated devices (e.
g.   chips) but respecting the inputs  and outputs  of the components.  This contrasts with the  softer   undirected degree sequence   invariant 
produced by preserving just the  number of  connections in  every node.   In this
case,  the relevance  relies on  the number  of relations  instead of
mattering   the   arrows  orientation -i.e., the undirected degree sequence. 
Furthermore,   the   conservation  of   the
\textit{connected components} is essential in a graph describing a \textit{process}, since fragmentation can be intepreted
as a break of the flow of causality.

Finally, we stress that an important feature of any randomization process is that topological invariants restrict the
space of graphical solutions. Our methodology provides valuable information about the \textit{randomness} of a particular structure within the
context of its graphical space of solutions. It is arguable to think that the higher the number of constraints the smaller the space of solutions.  In any case, its complete exploration is not feasible beyond a graph containing more than a handful of nodes. In this context, our methodology provides a sampling of such space in order to estimate the randomness of a DAG given some topological contraints


\section{Acknowledgements}
This  work  was  supported   by  the  EU  $6^{th}$  framework  project
ComplexDis (NEST-043241, CRC and JG), the UTE project CIMA (JG), James
McDonnell Foundation  (BCM and  RVS) and Santa  Fe Institute  (RVS) We
thank Complex System Lab members for fruitful conversations.




\end{document}